**Experimental Realization of Type-II Quadrupole Topological Insulator**

Yuan Liu,[1] Yangkai Liu,[1] Cheng Lin,[1] Jiao Shen,[1] Yifan Zhu,[1,2] Haiyan Fan,[1, 2, †] Hui Zhang[1, 2, †]

[1] Jiangsu Key Laboratory for Design and Manufacturing of Precision Medicine Equipment, School of Mechanical Engineering Southeast University, Nanjing, China.
[2] Key Laboratory of Underwater Acoustic Signal Processing (Southeast University), Ministry of Education, Nanjing China.

**ABSTRACT**. The discovery of quadrupole topological insulators (QTIs) has spurred extensive research into higher-order topological phases. Recently proposed type-II QTIs exhibit unconventional topological behaviors with 1/2 edge polarization $p_x$ and zero edge polarization $p_y$, due to the inequivalence between Wannier-band and edge-spectrum gap closures, yet their experimental realization remains challenging owing to the long-range and complex off-site hopping terms in their tight-binding model (TBM). Here, we circumvent this difficulty via an optimized Householder tridiagonalization (OHT) mapping that reduces the complex two-dimensional lattices to one-dimensional chains with only negative-real-valued nearest-neighbor hopping terms, greatly facilitating experimental sample fabrication. Using this strategy, we experimentally verify the type-II QTI phase, type-I QTI phase and trivial phase in elastic wave platforms via simple aperiodic plate-beam chain structures, where the plates reflect the on-site potential terms and beams correspond to the off-site hopping terms in the TBM. Our approach provides a versatile route for experimentally exploring more complex and richer topological phenomena based on TBM.

## I. INTRODUCTION

Topological materials exhibit quantized invariants that protect wave propagation against scattering and disorder, thereby endowing photonic and phononic devices with unprecedented robustness [1–4]. The advent of higher-order topological insulators (HOTIs) marks a profound extension of topological band theory from conventional first-order topological phases to higher-order topological phases [5–8]. Subsequent studies further demonstrated that the realization of such higher-order topological phases relies not only on specific tight-binding lattice designs but also on the identification of observable, quantized topological invariants [5,9–11], i.e., the quantized multipole moments including quadrupole [8,12], octupole [13,14], and hexadecapole moments [2,15]. Among them, quantized quadrupole moment [12,16], initially proposed in the Benalcazar-Bernevig-Hughes (BBH) model, is a typical topological invariant describing HOTIs. Despite these substantial theoretical works, considerable experimental efforts have also been devoted in HOTIs by diverse classical wave platforms including photonic [17,18], phononic [19,20], elastic wave [21], microwave [22–24] and electrical circuits [25]. Nevertheless, the topological classification of higher-order phases remains incomplete by conventional multipole paradigm. Meanwhile, theoretical studies have revealed that the closure of the Wannier gap does not always occur simultaneously with that of the edge band gap [26], leading to the classification of quadrupole topological insulators (QTIs) into four types: type-I [14,15], type-I anomalous, type-II, and type-II anomalous [26]. For type-II QTIs, edge polarizations exist only on a single pair of boundaries, breaking the correspondence dictated by classical electric multipole theory and offering a new perspective on the diversity of higher-order topological phases. The key lies in introducing long-range couplings to realize novel edge polarizations while preserving reflection and particle-hole symmetries [26]. However, the experimental realization of type-II QTIs faces substantial challenges. First, the design of long-range hoppings involves off-site terms and phases across multiple sites, whose signs and strengths must be precisely controlled to maintain the symmetries, leading to considerable complexity in the tight-binding model (TBM) [27–29]. Second, realizing this topological phase often requires a sufficiently large lattice size [26] to ensure the clear resolution of zero-energy corner-localized states and the accurate characterization of their topological properties, which further increases the difficulty of sample fabrication and measurement.

Recent works have validated the Lanczos method [30] for four-dimensional topological insulators with nearest-neighbor coupling and the Householder method [31] for Z-classified phases with simple long-range coupling. The simplified methods outlined above nonetheless point to promising strategies for tackling the experimental challenges of such systems. However, the methods are not directly suitable for simplifying type-II QTIs with complex long-range hoppings and large system sizes.

In this work, by employing a uniquely optimized Householder tridiagonalization (OHT), the two-dimensional (2D) type-II QTI Hamiltonian—characterized by complex long-range hoppings and edge polarizations—is rigorously mapped onto a 1D aperiodic disorder chain with uneven on-site potentials and off-site hopping terms. The resulting tridiagonal Hamiltonian matrix, approximately isospectral with the original, faithfully preserves key physical features such as corner states. Through identification of localized corner states and near-zero off-site amplitudes, the resulting large-sized tridiagonal Hamiltonian matrix are truncated, leading to chains with 24, 17, and 8 sites for the type-II QTI phase, type-I QTI phase, and trivial phase, respectively. Subsequently, the chains are mapped onto plate-beam structures, where the on-site potentials correspond to the plate resonant frequencies, and the off-site hopping terms correspond to the coupling frequencies induced by the thin beams. For experimental validation, the mode shape of the single plate and the intensity distributions of the truncated elastic chains are obtained. The good agreement among TBM predictions, finite-element simulations, and experimental measurements confirms the robustness of this mapping approach across all three topological phases. This approach simplifies complex topological systems, reduces complexity of the geometric structures, and enhances experimental feasibility, thereby addressing the scarcity of experimental investigations into type-II QTI and providing a scalable platform for exploring complex topological states.

## II. RESULTS

### A. Type-II QTI model

The 2D system of interest is a square lattice of 4-site unit cells with complex long-range couplings, as shown in Fig. 1(a). The orange/green/blue/yellow solid lines represent the nearest-neighbor hoppings inside a unit cell/the long-range hoppings to the nearest-neighbor unit cells along the x-axis direction/the long-range hoppings to the nearest-neighbor unit cells along the $y$-axis direction/ the long-range hoppings to the next-nearest-neighbor unit cells along the y-axis direction, denoted as $h_{(00)}$ / $h_{(10)}$ / $h_{(01)}$ / $h_{(02)}$. The orange, green, blue, yellow dashed lines in Fig. 1(a) are the corresponding negative hoppings, denoted as $-h_{(00)}$, $-h_{(10)}$, $-h_{(01)}$, $-h_{(02)}$, respectively. The corresponding tight-binding Hamiltonian can be described as

$$H_{II} = H_L + \sigma_2 \otimes H'_{SSH}(k_x). \quad (1)$$

Here,

$$H_L = (\gamma + \cos k_y)\sigma_1 \otimes \sigma_0 + (\sin k_y + b_2 \sin 2k_y)\sigma_3 \otimes \sigma_1 + b_2 \cos 2k_y \sigma_3 \otimes \sigma_2, \quad (2)$$

describes the hopping situation in the $y$-direction. In Eq. (2), $\sigma_0$, $\sigma_1$, $\sigma_2$ and $\sigma_3$ are all Pauli matrices, k is the Bloch wave number, $\gamma$ is the nearest-neighbor intercell hopping amplitude and $b_2 = 0.8$ is the long-range hopping amplitude in the $y$-direction.

$$H'_{SSH}(k_x) = (\gamma + t_x - g_0 + t_x \cos k_x)\sigma_2 + t_x \sin k_x \sigma_3, \quad (3)$$

describes the hopping situation in the x-direction and can be obtained by applying a unitary transformation to $H_{SSH}(k_x)$. Here, $t_x = 1$ represents the nearest-neighbor intercell hopping amplitude in the $y$-direction, $g_0 = 0.22$ is a constant offset and $\otimes$ denotes the Kronecker product. The emergence of quantified multipole moments and corner states in type-II QTIs is closely related to long-range hoppings, the reflection symmetries ($\hat{m}_x = \sigma_1 \otimes \sigma_3$, $\hat{m}_y = \sigma_1 \otimes \sigma_1$) and particle-hole symmetry ($\Xi = \sigma_3 \otimes \sigma_0 \kappa$, $\kappa$ is the complex conjugation operator).

The eigen-energy evolution of the lattice composed of $9 \times 9$ unit cells with increasing $\gamma$ can be found in Fig. 1(b), where the $\alpha$ (red dots) represent corner states, the $\beta$ (blue dots) represent edge states, and the gray dots represent bulk states, the identification method can be found in Supplemental Material [32]. By increasing the value of $\gamma$, two topological phase transitions occur, from trivial to type-I QTI and then to type-II QTI. The Wannier spectra for three distinct topological phases [marked by the vertical lines in Fig. 1(b)] are denoted in Figs. 1(c)-1(e), respectively, displaying $\nu_x(\nu_y)$ with periodic boundary conditions along $x(y)$ and open boundary conditions along $y(x)$, where isolated Wannier centers are highlighted by red dots. The selected eigenmode profiles of trivial, type-I QTI, and type-II QTI are shown in Figs. 1(f)-1(h), respectively. The bulk, edge and corner states can be clearly observed in type-I QTI [Fig. 1(g)] and type-II QTI [Fig. 1(h)] phases, but only bulk states can be observed in trivial phase [Fig. 1(f)]. Note that the edge states in type-II QTI only exhibit localization in the x direction, while edge states both in *x* and *y* directions can be observed in type-I QTI. Because the edge states in type-I QTI satisfy $p_y = p_x = 0.5$, while those in type-II QTI satisfy $p_y = 0$ and $p_x = 0.5$, revealing the essential difference between the two types of QTIs [26]. As shown in Fig. 1(a), the aforementioned Type-II QTI model can be mapped onto an aperiodic chain with only nearest-neighbor hopping via the OHT method, which significantly simplifies the TBM of the type-II QTI.

### B. The optimized Householder tridiagonalization (OHT)

The conventional Householder algorithm cannot transform a complex matrix into a purely positive-real-valued matrix, which increases the complexity of subsequent experimental samples. Therefore, we optimize the Householder algorithm to ensure that the transformed matrix only possess positive real numbers. By means of phase rotations, we can maintain good numerical stability and more detailed discussion is shown in Supplemental Material [32]. After applying OHT to a finite 2D type-II QTI, an aperiodic chain with disordered off-site hopping amplitudes and on-site potential terms can be obtained. The eigen-energy spectrum of the transformed tridiagonal Hamiltonian matrix [red triangles in Fig. 2(a)] shows good consistence with the spectrum of the original matrix [black dots in Fig.2(a)].

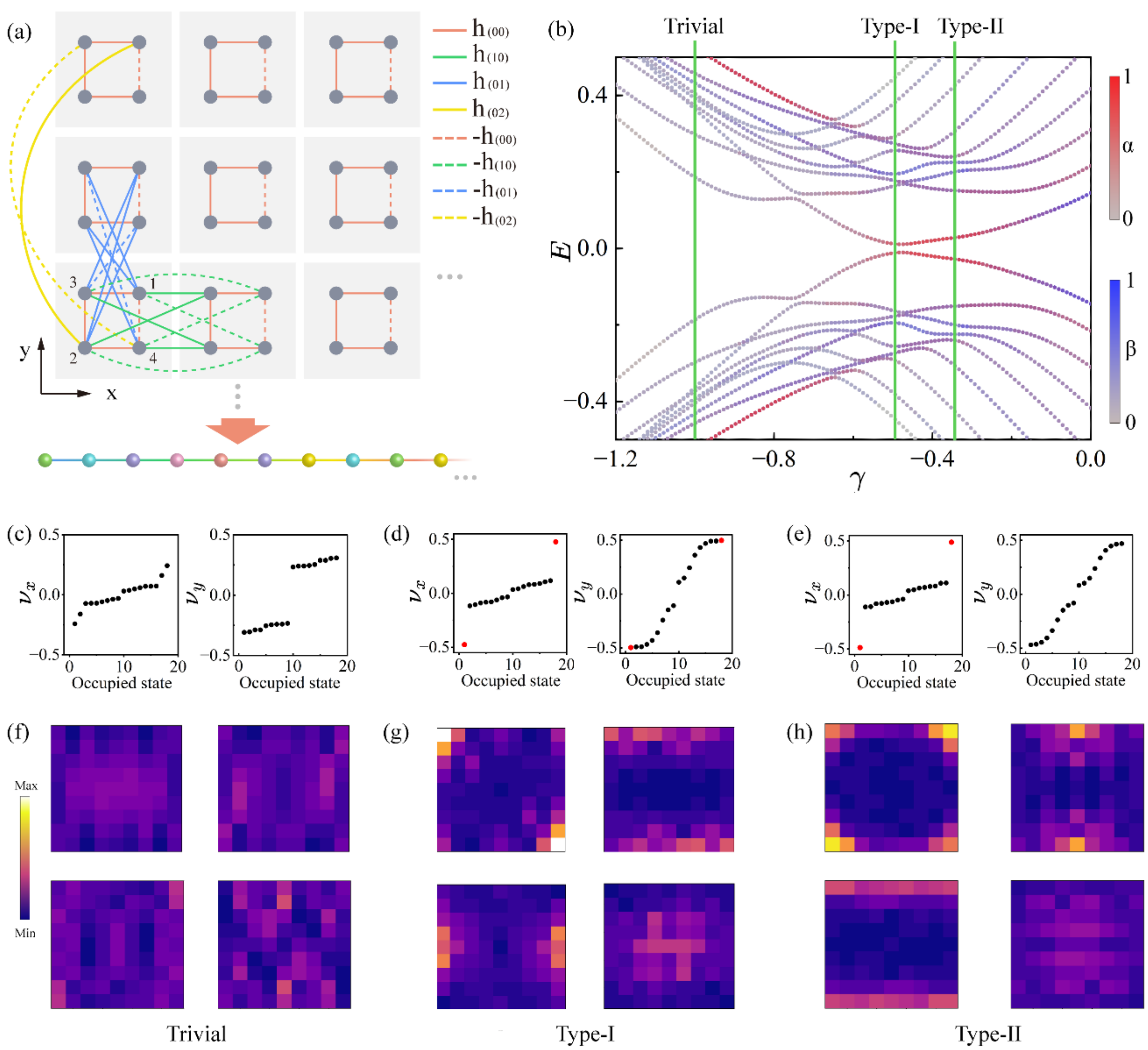

FIG 1. The TBM calculation about the QTI model. (a) The schematic diagram of the original TBM with type-II QTI. The bottom of the graph is the schematic diagram of the transformed TBM with aperiodic on-site potential and off-site hoppings. (b) The eigen-energy spectrum as a function of $\gamma$. The red, blue, and gray dots indicate the corner, edge, and bulk states, respectively. The $\gamma$ parameters for the three topological phases – trivial $(\gamma = -1)$, type-I QTI $(\gamma = -0.5)$, and type-II QTI $(\gamma = -0.35)$– are marked by the green vertical solid lines. (c), (d), (e) Wannier spectra corresponding to the trivial, type-I QTI, and type-II QTI phases, respectively. The isolated Wannier centers are highlighted by red circles. (f), (g), (h) Eigenmodes of trivial, type-I QTI, and type-II QTI, respectively. Each pixel denotes the sum of eigenmode probabilities over one unit cell.

For the transformed 1D disordered chain with only nearest-neighbor hoppings, the Hamiltonian is given by

$$H_{1D} = V^{'} H_{II} V = \sum_{n=1}^{N} a_n c_n^{\dagger} c_n + \sum_{n=1}^{N-1} b_n (c_n^{\dagger} c_{n+1} + c_{n+1}^{\dagger} c_n), \quad (4)$$

where V is the Householder transformation operator, $c_n(c_n^{\dagger})$ represents the creation (annihilation) operator on the $n$ site, and $a_n\,(b_n)$ is the effective on-site potential (off-site hopping amplitude) on the $n$ site of the 1D array. During the OHT, we select the first element of the matrix as the anchor site by default. After the OHT, the on-site and off-site terms of the 1D chain with type-II QTI $(\gamma = -0.35)$ are shown in Fig. 2(c).

As shown in the subgraph of Fig. 2(a), there are 4 zero-energy modes. The spatial distribution probabilities of the four zero-energy modes in 1D chain are shown in Fig. 2(b). This chain contains a corner state with high probability at site 193, while the other three corner states become more dispersed after the transformation. Since the 1D 324-site chain is too long for experiments, we need to truncate it. Two requirements need to be satisfied during truncation: (1) cutting at positions where hopping terms $b_n \approx 0$, so that the influence on the chain properties is minimized, (2) including site 193 with high probability, as this site holds the target corner state to be investigated in experiments. Accordingly, $n_1 = 192$ and $n_2 = 216$ are selected to be the truncation positions [Fig.2 (c)]. Subsequent experiments will only be conducted on the truncated chain with 24 sites.

The ratio of the maximum to the minimum of the shifted on-site potentials is expressed as

$$S = (a_{n\,\max} + C_{II})/(a_{n\,\min} + C_{II}). \quad (5)$$

The maximum (minimum) value of on-site potentials in the truncated chain is $a_{n\,\max}$ ( $a_{n\,\min}$ ). $C_{II} = 49.831$ is a constant shift on the on-site potentials of the 1D chain to maintain effective coupling between two sites in the following experimental process. In the present case, $S = 1.05$. The on-site and off-site terms of the truncated type-II QTI chain after applying the constant shift is shown in Fig. 2(d). Note that the energy spectrum and probability distribution is unchanged after the constant shift.

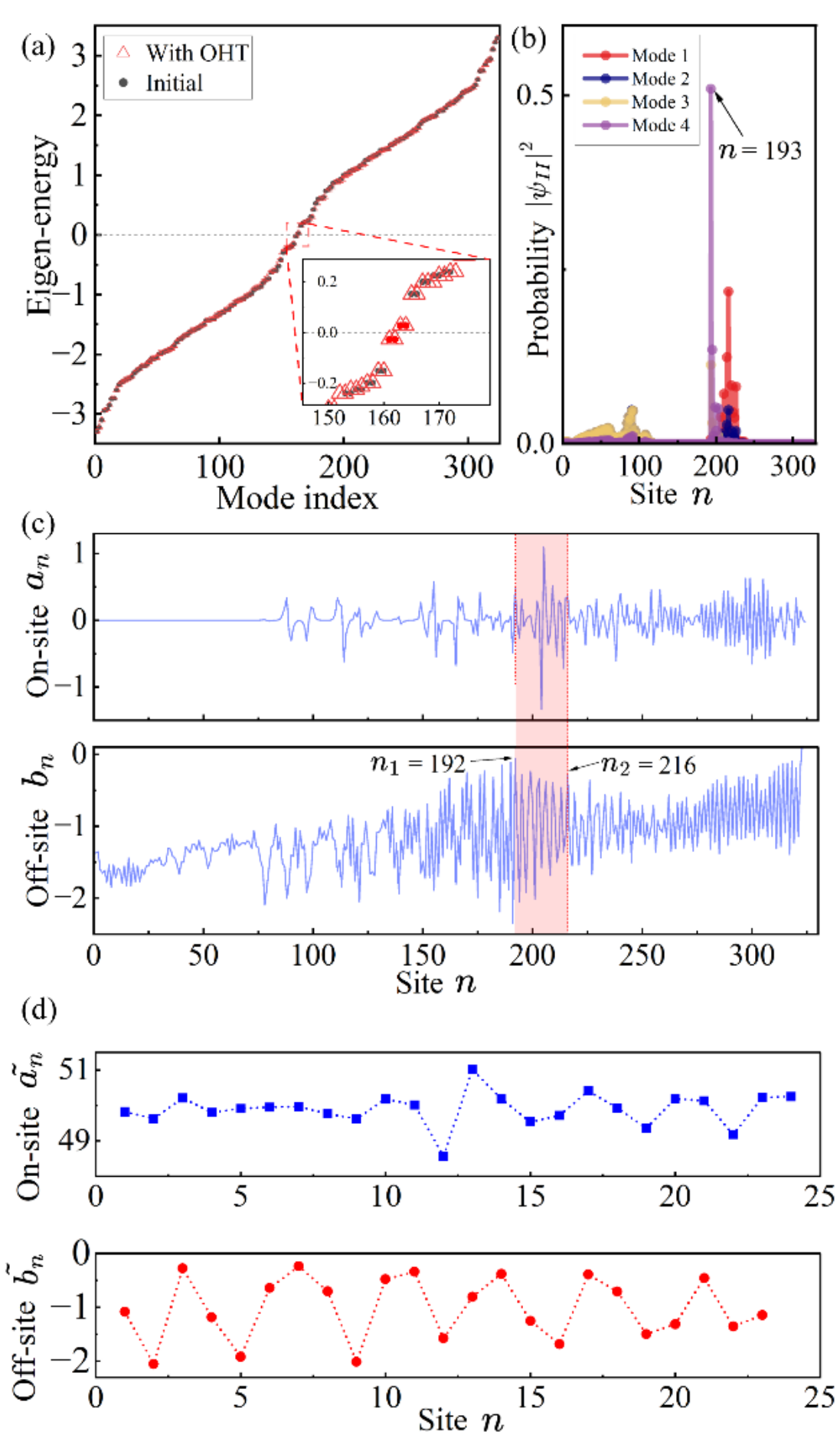

FIG 2. The tight binding results of the transformed Hamiltonian of the type-II QTI. (a) The comparison of the eigen-energy spectra for a finite (9×9×4 sites) type-II QTI lattice and the aperiodic 1D chain. The black circles represent the initial energy spectrum, while the red triangles represent the energy spectrum with OHT. The subgraph is the enlarged view around zero-energy, showing 4 zero-energy modes. (b) The four corner wavefunctions of the 1D chain, with a highly localized corner state at site 193. (c) On-site and off-site terms of the 1D disordered chain. The red area represents the truncated part, which means the chain is truncated at the coupling indices $n_1 = 192$ and $n_2 = 216$. (d) The on-site amplitudes $\widetilde{a_n}$ and off-site hoppings $\widetilde{b_n}$ of the truncated chain after applying the constant shift $C_{II}$.

Applying the transformation matrix V[Eq. (4)] to the 1D chain can map it back to 2D configuration [30], yielding the mode diagram of the inverse transformation. Specific information of the mode is shown in Supplemental Material [32]. Comparison of the original and inversely transformed eigenmodes diagrams validates that the 1D truncated chain effectively preserves the essential physical information of the original 2D model.

## III. Elastic realization of type-II QTI

In this section, we perform dimensionality reduction and truncation on the three topological phases, forming experimental samples that are verified separately.

### *1. Experiments for the type-II QTI*

As shown in Fig. 2(d), a truncated chain containing 24 sites, which is suitable for experiments. The above TBM can be realized using elastic resonant plates connected by thin beams, as depicted in Fig. 3(e), where the on-site potential terms correspond to the resonant frequency of the plates, and off-site hopping terms correspond to coupled frequencies endowed by the thin beams. The coupling strength $|\kappa| = |f_2 - f_1|/2$, where $f_1$, $f_2$ are the resonant frequencies of the coupled two plates structures stem from the first non-rigid-body resonant mode of a single plate [21]. The coupling strengths are closely related to the geometric parameters of the connecting beams, such as width $w$, position $h$, and length $d$ [21]. Specific information of the geometric data is shown in Supplemental Material [32]. The eigenenergies obtained by TBM [Fig. 3(a)] and eigenfrequencies obtained by numerical simulations [Fig. 3(b)] show good agreement, where the red (black) dots denote the corner (bulk) states. This confirms that the finite plate-beam model meets the requirement of the TBM. The wavefunction distributions of the truncated chain are obtained from the TBM, showing a corner state localized at the first site [Fig. 3c].

For the plate-beam system, the unavoidable inherent damping is represented by the imaginary component of the Young's modulus, with a loss factor of $\eta = 0.005$ obtained by matching the simulated frequency response spectrum [red line in Fig.3(d)] with the measured ones [blue circles in Fig. 3(d)]. The plate-beam structure is cut from an aluminum alloy plate (Young's modulus $E = 71.2$ GPa, density $\rho = 2700\text{kg/m}^3$, Poisson's ratio $\nu = 0.33$) by laser. In the experiment, both ends of the sample are clamped to mimic fixed boundary conditions. The sample is excited by a shaker (ZHENYAN TECH PME50), and the out-of-plane vibrations of each plate are collected by a laser vibrometer (HUAQIN OPTACOUS SLV-1550), yielding the corresponding amplitude signals of the plates.

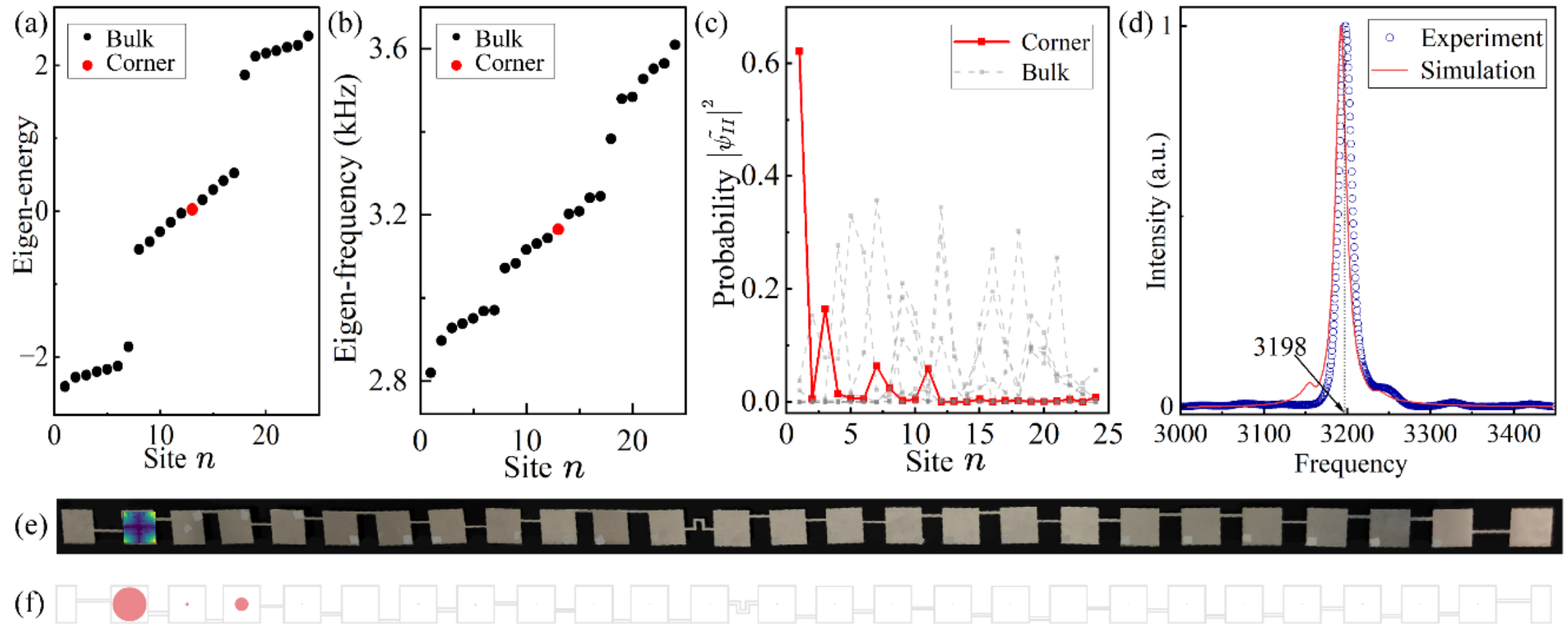

FIG 3. Simulation and experimental diagrams of the type-II QTI truncated chain. (a) Eigen-energy spectrum of the truncated chain containing 24 sites calculated from the TBM. The red dot indicates the topological corner states at zero-energy. (b) Eigen-frequencies of the plate-beam elastic structure obtained from finite-element simulations. The red dot indicates the topological corner state. (c)The intensity distributions of the truncated elastic chain based on the TBM, where the corner distribution is marked by the red solid line, bulk distributions are denoted by the gray dashed

line. (d) Frequency response results from 3000 Hz to 3450 Hz. The blue circles (red line) represent the experimental (simulated) results. The resonant peak is located at 3198 Hz. (e) Experimental sample fabricated by laser-cut technic from an aluminum alloy plate. The subgraph on the first plate is the measured mode shape of a single square plate. (f) The spatial distribution of the corner states measured at 3198 Hz. The sizes of the red circles on each plate represent the intensity.

First, we excite the first plate and detect its response in the same plate while sweeping the frequency from 3000 Hz to 3450 Hz with the resolution of 0.5 Hz. The measured frequency response shows a peak at 3198 Hz [blue circles in Fig. 3(d)], which is consistent with the eigenfrequency of the corner states in Fig. 3(b). Then, the excitation frequency is pinned at 3198 Hz, and the vibration amplitude on each plate is measured. The resulting spatial distribution [Fig. 3(f)] shows good agreement with the TBM results presented by the red line in Fig. 3(c).

### *2. Experiments for the type-I QTI and trivial phases*

Similarly, we apply the above truncation method to the type-I QTI (trivial) system with $\gamma = -0.5(\gamma = -1)$ to verify the evolution of the topological phase with $\gamma$. First, the OHT is performed on the type-I QTI Hamiltonian [Eq. (1)-(3)]. As shown in the spatial distributions of the four corner states [Fig. 4(a)], one of them is highly localized at site 191. Hence, the truncated chain should contain this site. As a result, the 1D chain with 324-site is truncated at site 198 and site 205 where coupling strength is -0.046 and -0.365, respectively. Then, a truncated chain consisting of 17 sites is obtained. Specific information of the truncation is shown in Supplemental Material [32]. As mentioned in Eq. (5), the constant shift ($C_I = 46.567$) is also applied to the on-site terms of the type-I QTI system to satisfy $S = 1.05$. Shown in Figs. 4(b) and 4(c) are the eigen-energy of the truncated chain and the eigen-frequency of the plate–beam structure, with red(black) dots denoting the corner(bulk) states.

Next, the intensity distributions of the truncated chain are calculated, showing a corner state localized at the third plate [red line in Fig. 4(d)]. The truncated chain is mapped onto the plate-beam model for numerical simulation and experimental validation [Fig. 4(f)].

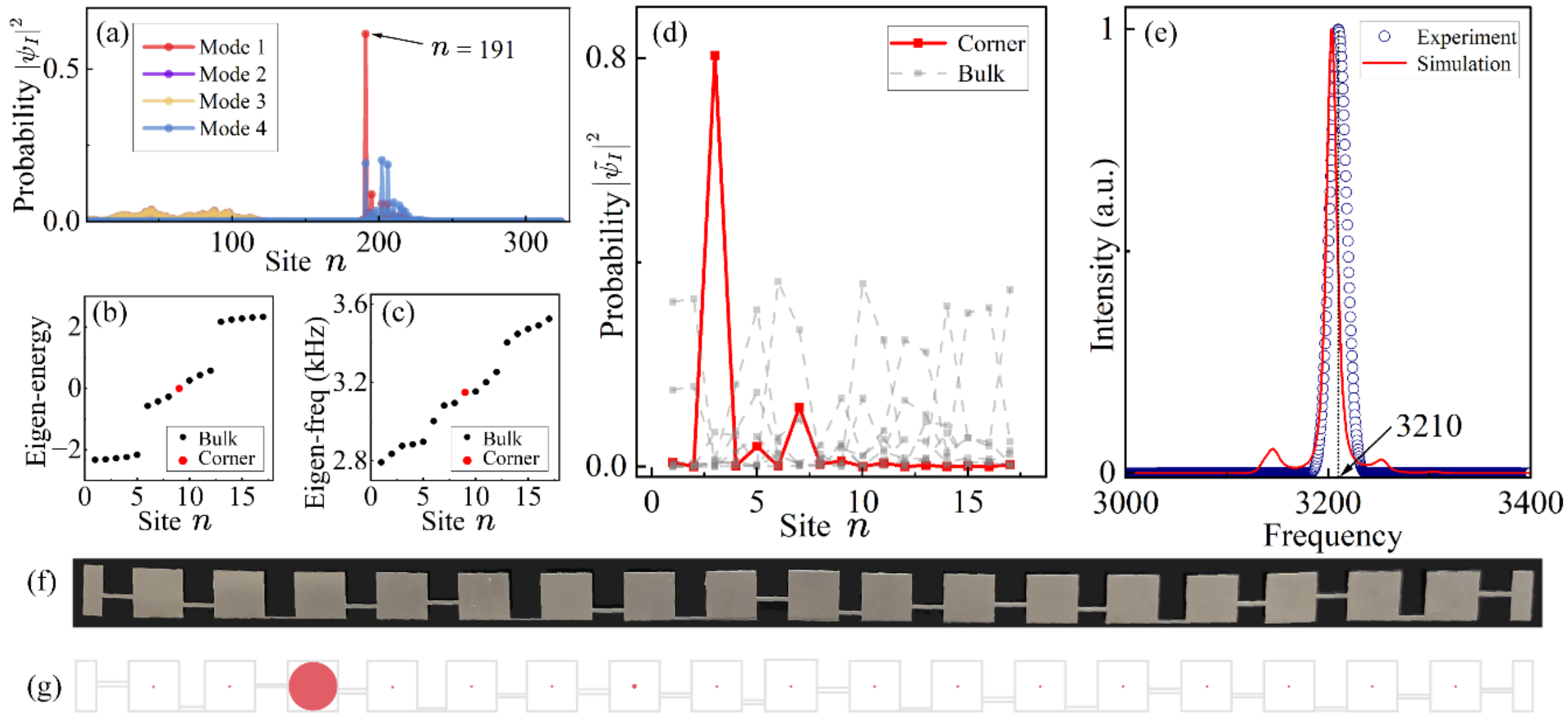

FIG 4. Simulated and experimental results of type-I QTI truncated chain. (a) The spatial distribution probabilities of the four corner states of 1D chain, with a highly localized corner state at site 191. (b) Eigen-energy spectrum of the truncated chain containing 17 sites calculated from the TBM. The red dot indicates the topological corner states at zero-energy. (c) Eigenfrequencies of the plate-beam acoustic structure, obtained via finite-element simulations, with the red(black) dot identifying the topological zero-energy(bulk) state. (d)Wavefunction distribution of the corner state in the truncated chain obtained from the TBM. The corner state is localized at the first site. (e) Frequency sweep results

from 3000 Hz to 3400 Hz. Experimental measurements reveal energy localization at 3210 Hz. (f) Experimental sample fabricated by laser-cut aluminum alloy. (g) Vibration amplitude response measured at an incident acoustic wave frequency of 3210 Hz. The radius of the red circle on each plate scales with the amplitude energy magnitude.

In the experiment, the third plate (the plate hosting the corner state) is excited, and the response is detected in the same plate while sweeping the frequency from 3000 Hz to 3400 Hz. The measured frequency response is centered at 3210 Hz [blue circles in Fig. 4(e)]. Then, the excitation is continued on the third plate, with the frequency set to the characteristic frequency of the corner state, i.e., 3210 Hz. The vibration amplitude on each plate at this excitation frequency is then measured. The resulting amplitude distribution [Fig. 4(g)] shows good agreement with the TBM results presented in Fig. 4(d).

Experimental assistance is provided in verifying the trivial Wannier band diagram in Fig. 1(e) and the trivial mode diagram in Fig. 1(h), demonstrating that no localized states exist in the trivial phase system. For the trivial phase, the system is mapped onto a 1D chain via the OHT. The 1D chain is truncated at a position where $n_1 = 192 (n_2 = 216)$ and coupling strength is $-0.153(-0.123)$, a truncated chain containing 8 sites is obtained. Then, a constant shift ($C_t = 21.873$) is applied to the original on-site energies to satisfy $S = 1.05$. Specific information of the truncation is shown in Supplemental Material [32]. Next, the energy spectrum and the corresponding wavefunction distribution of the truncated chain are shown in Figs. 5(a) and 5(c), respectively. For simulation and testing, this chain is mapped onto the plate-beam model depicted in Fig. 5(e). The resulting frequency spectrum of the model is presented in Fig. 5(b), which confirms the TBM results in Fig. 5(a). In experiment, the bulk state is the target to test. The third plate is excited, and the frequency-domain response is detected on the fourth plate from 2600 Hz to 3600 Hz. Frequency peaks corresponding to bulk states are observed at 2717.5 Hz and 3535.5 Hz [blue circles in Fig. 5(d)]. Subsequently, with the excitation frequency set to 2717.5 Hz on the third plate, the vibration amplitude on each plate is measured at this frequency [Fig. 5(f)]. The results show that the energy is mainly concentrated on the second and third plates [Fig. 5(c)].

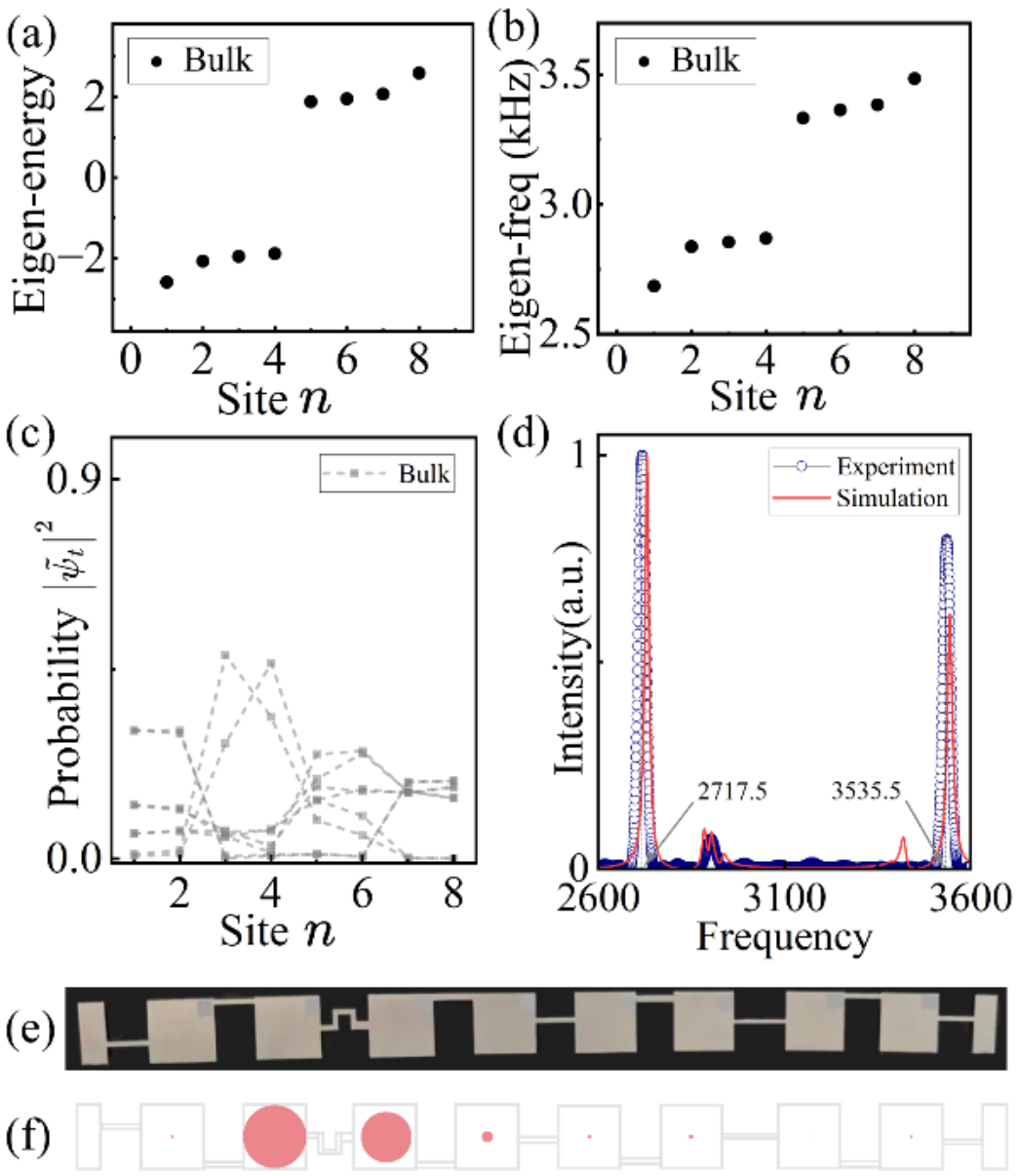

FIG 5. Simulation and experimental results of trivial phase truncated chain. (a) Eigen-energy spectrum of the truncated chain containing 8 sites calculated from the TBM. (b) Eigen-frequencies of the plate-beam acoustic structure obtained from finite-element simulations. (c) The wavefunction distribution of the corner state in the truncated chain was derived from TBM. (d) Frequency sweep from 2600 Hz to 3600 Hz, with bulk-state peaks identified at 2717.5 Hz and 3535.5 Hz. (e) Experimental sample fabricated by laser-cut aluminum alloy. (f) Vibration amplitude response measured at an incident acoustic wave frequency of 2717.5 Hz. The radius of the red circle on each plate is proportional to the corresponding amplitude energy magnitude.

## III. DISCUSSION & CONCLUSION

In summary, we have proposed and experimentally demonstrated a dimensionality reduction strategy for the verification of type-II QTI in elastic wave systems. The OHT combined with phase rotation can change the complex matrix with long-range hoppings and different phases into a pure negative-real matrix with only nearest-neighbor hoppings, thus significantly reducing the geometrical complexity of experimental samples. By employing the OHT, the 2D type-II QTI Hamiltonian — characterized by long-range couplings and edge polarizations — is rigorously mapped onto a 1D aperiodic disorder chain with uneven on-site potentials and off-site hoppings. The transformed tridiagonal Hamiltonian matrix, which is approximately isospectral with the original system, strictly preserves the physical features of the initial 2D Hamiltonian, such as corner states. By identifying the highly localized corner states and confirming the near-zero hopping amplitude, the truncated chains composed of 24, 17, and 8 sites for the type-II, type-I, and trivial phases are extracted, respectively. The three truncated chains are all mapped onto the plate-beam structures for experimental investigations. The good agreement between the TBM predictions, finite-element simulations, and experimental measurements validates the robustness of the proposed mapping method across all three topological phases (type-II, type-I and trivial phases). This method can be readily extended to other classical wave systems

This OHT dimensional reduction strategy not only retains the corner-state wavefunction with high localization fidelity, but also dramatically reduces the system size required for experimental realization. This method circumvents the fundamental challenge that type-II QTIs typically demand large lattices and complex couplings for protecting the corner states. More broadly, the present approach establishes a universal and scalable route for experimentally accessing complex topological models that are otherwise prohibitively difficult to realize directly. The method is not restricted to type-II QTIs: any Hermitian tight-binding Hamiltonian — regardless of dimensionality, range of couplings, or complexity of the unit cell — can in principle be reduced to a 1D aperiodic nearest-neighbor chain via the same procedure. Furthermore, the demonstrated high locality of the corner states provides a property of practical relevance for device applications such as small size topological mechanical sensors.

## ACKNOWLEDGMENTS

This work was supported by the National Natural Science Foundation of China (Grant No.52501418) and the Natural Science Foundation of Jiangsu Province (Grant No. BK20251266).

[1] Y.-F. Chen, Z.-G. Chen, H. Ge, C. He, X. Li, M.-H. Lu, X.-C. Sun, S.-Y. Yu, and X. Zhang, Various topological phases and their abnormal effects of topological acoustic metamaterials, Interdisciplinary Materials 2, 179 (2023).

[2] A. Dutt, M. Minkov, I. A. D. Williamson, and S. Fan, Higher-order topological insulators in synthetic dimensions, Light Sci Appl 9, 131 (2020).

[3] X. Zhang, M. Xiao, Y. Cheng, M.-H. Lu, and J. Christensen, Topological sound, Commun Phys 1, 97 (2018).

[4] G. Ma, M. Xiao, and C. T. Chan, Topological phases in acoustic and mechanical systems, Nat Rev Phys 1, 281 (2019).

[5] M. Ezawa, Higher-Order Topological Insulators and Semimetals on the Breathing Kagome and Pyrochlore Lattices, Phys. Rev. Lett. 120, 026801 (2018).

[6] B.-Y. Xie, H.-F. Wang, H.-X. Wang, X.-Y. Zhu, J.-H. Jiang, M.-H. Lu, and Y.-F. Chen, Second-order photonic topological insulator with corner states, Phys. Rev. B 98, 205147 (2018).

[7] W. A. Benalcazar, T. Li, and T. L. Hughes, Quantization of fractional corner charge in ${C}_{n}$-symmetric higher-order topological crystalline insulators, Phys. Rev. B 99, 245151 (2019).

[8] M. Serra-Garcia, V. Peri, R. Süsstrunk, O. R. Bilal, T. Larsen, L. G. Villanueva, and S. D. Huber, Observation of a phononic quadrupole topological insulator, Nature 555, 342 (2018).

[9] S. Ren, I. Souza, and D. Vanderbilt, Quadrupole moments, edge polarizations, and corner charges in the Wannier representation, Phys. Rev. B 103, 035147 (2021).

[10] Y. Wang et al., Quantum superposition demonstrated higher-order topological bound states in the continuum, Light Sci Appl 10, 173 (2021).

[11] Z. Du, J. Luo, Z. Xu, Z. Jiang, X. Ding, T. Cui, and X. Guo, Higher-order topological insulators by ML-enhanced topology optimization, International Journal of Mechanical Sciences 255, 108441 (2023).

[12] W. A. Benalcazar, B. A. Bernevig, and T. L. Hughes, Quantized electric multipole insulators, Science 357, 61 (2017).

[13] H. Xue, Y. Ge, H.-X. Sun, Q. Wang, D. Jia, Y.-J. Guan, S.-Q. Yuan, Y. Chong, and B. Zhang, Observation of an acoustic octupole topological insulator, Nat Commun 11, 2442 (2020).

[14] X. Ni, M. Li, M. Weiner, A. Alù, and A. B. Khanikaev, Demonstration of a quantized acoustic octupole topological insulator, Nat Commun 11, 2108 (2020).

[15] W. Zhang, D. Zou, J. Bao, W. He, Q. Pei, H. Sun, and X. Zhang, Topolectrical-circuit realization of a four-dimensional hexadecapole insulator, Phys. Rev. B 102, 100102 (2020).

[16] Benalcazar, Wladimir A., Bernevig, B. Andrei, and Hughes, Taylor L., Electric multipole moments, topological multipole moment pumping, and chiral hinge states in crystalline insulators | Phys. Rev. B, Phys. Rev. B 96, 245115 (2017).

[17] F. Zanetto, F. Toso, M. Crico, F. Morichetti, A. Melloni, G. Ferrari, and M. Sampietro, Unconventional Monolithic Electronics in a Conventional Silicon Photonics Platform, IEEE Transactions on Electron Devices 70, 4993 (2023).

[18] T.-C. Hsueh, Y. Fainman, and B. Lin, Monolithic Silicon-Photonics Linear-Algebra Accelerators Enabling Next-Gen Massive MIMO, Journal of Lightwave Technology 42, 7903 (2024).

[19] S. Qu, W. Ding, L. Dong, J. Zhu, S. Zhu, Y. Yang, and W. Zhai, Chiral phononic crystal-inspired railway track for low-frequency vibration suppression, International Journal of Mechanical Sciences 274, 109275 (2024).

[20] Y. Chen, W. Qin, S. Zhang, P. Cui, Q. Niu, and Z. Zhang, Emergence of Chiral Phonons in Two-Dimensional Kagome Lattices Harboring Electronic Chirality, Phys. Rev. Lett. 135, 126608 (2025).

[21] H. Fan, H. Gao, S. An, Z. Gu, S. Liang, Y. Zheng, and T. Liu, Hermitian and non-hermitian topological edge states in one-dimensional perturbative elastic metamaterials, Mechanical Systems and Signal Processing **169**, 108774 (2022).

[22] C. W. Peterson, W. A. Benalcazar, T. L. Hughes, and G. Bahl, A quantized microwave quadrupole insulator with topologically protected corner states, Nature **555**, 346 (2018).

[23] P. Harvey-Collard, J. Dijkema, G. Zheng, A. Sammak, G. Scappucci, and L. M. K. Vandersypen, Coherent Spin-Spin Coupling Mediated by Virtual Microwave Photons, Phys. Rev. X **12**, 021026 (2022).

[24] X. Han, W. Fu, C.-L. Zou, L. Jiang, and H. X. Tang, Microwave-optical quantum frequency conversion, Optica, OPTICA **8**, 1050 (2021).

[25] S. Imhof et al., Topolectrical-circuit realization of topological corner modes, Nature Phys **14**, 925 (2018).

[26] Y.-B. Yang, K. Li, L.-M. Duan, and Y. Xu, Type-II quadrupole topological insulators, Phys. Rev. Research **2**, 033029 (2020).

[27] Wang, Dongyi, Deng, Yuanchen, Ji, Jun, Oudich, Mourad, Benalcazar, Wladimir A., Ma, Guancong, and Jing, Yun, Realization of a Z-Classified Chiral-Symmetric Higher-Order Topological Insulator in a Coupling-Inverted Acoustic Crystal, Physical Review Letters **131**, 157201 (2023).

[28] Y. Yang, Z. Gao, H. Xue, L. Zhang, M. He, Z. Yang, R. Singh, Y. Chong, B. Zhang, and H. Chen, Realization of a three-dimensional photonic topological insulator, Nature **565**, 622 (2019).

[29] M. Weiner, X. Ni, M. Li, A. Alù, and A. B. Khanikaev, Demonstration of a third-order hierarchy of topological states in a three-dimensional acoustic metamaterial | Science Advances, Sci. Adv. (Science Advances) **6**, eaay4166 (2020).

[30] K. Chen, M. Weiner, M. Li, X. Ni, A. Alù, and A. B. Khanikaev, Nonlocal topological insulators: Deterministic aperiodic arrays supporting localized topological states protected by nonlocal symmetries, Proceedings of the National Academy of Sciences **118**, e2100691118 (2021).

[31] Y. Sun, L. Wang, H. Duan, and J. Wang, A new class of higher-order topological insulators that localize energy at arbitrary multiple sites, Science Bulletin **70**, 667 (2025).

[32] See Supplemental Material at Link for the delineation of corner, edge and bulk states, the details of OHT, the truncation positions for the type-I and trivial phases and the specific geometric data and experimental details of plate-beam structure.

.